\documentclass{emulateapj}

\shorttitle{HD 98618: A Star Similar to our Sun}
\shortauthors{Mel\'endez, Dodds-Eden, Robles}

\begin{document}
\title {HD 98618: A Star Closely Resembling our Sun\altaffilmark{1}}

\newcommand{\teff}{T$_{\rm eff}$ }
\newcommand{\tsin}{T$_{\rm eff}$}
\newbox\grsign \setbox\grsign=\hbox{$>$} \newdimen\grdimen \grdimen=\ht\grsign
\newbox\simlessbox \newbox\simgreatbox
\setbox\simgreatbox=\hbox{\raise.5ex\hbox{$>$}\llap
     {\lower.5ex\hbox{$\sim$}}}\ht1=\grdimen\dp1=0pt
\setbox\simlessbox=\hbox{\raise.5ex\hbox{$<$}\llap
     {\lower.5ex\hbox{$\sim$}}}\ht2=\grdimen\dp2=0pt\def\simgreat{\mathrel{\copy\simgreatbox}}
\def\simless{\mathrel{\copy\simlessbox}}
\newbox\simppropto
\setbox\simppropto=\hbox{\raise.5ex\hbox{$\sim$}\llap
     {\lower.5ex\hbox{$\propto$}}}\ht2=\grdimen\dp2=0pt
\def\simpropto{\mathrel{\copy\simppropto}}

\author{Jorge Mel\'endez\altaffilmark{2}, Katie Dodds-Eden \& Jos\'e A. Robles} 
\affil{Research School of Astronomy \& Astrophysics, 
Mt. Stromlo Observatory, Cotter Rd., Weston Creek, ACT 2611, Australia} 
\email{jorge,katie,josan@mso.anu.edu.au}

\altaffiltext{1}{The data presented herein were obtained at the W.M. Keck Observatory, 
which is operated as a scientific partnership among the California Institute
of Technology, the University of California and the NASA.}

\altaffiltext{2}{Also Department of Astronomy, 
California Institute of Technology; and affiliated with the Seminario Permanente 
de Astronom\'ia y Ciencias Espaciales of the 
Universidad Nacional Mayor de San Marcos, Peru.}

\slugcomment{Submitted to the The Astrophysical Journal Letters}
\slugcomment{Send proofs to:  J. Melendez}

\begin{abstract}

Despite the observational effort carried out in the last few decades,
no perfect solar twin has been found to date. An important milestone was
achieved a decade ago by Porto de Mello \& da Silva, who showed
that 18 Sco is almost a solar twin. In the present work,
we use extremely high resolution (R = 10$^5$) high S/N Keck HIRES 
spectra to carry out a differential analysis of 16 solar twin candidates. 
We show that HD 98618 is the second-closest solar twin
and that the fundamental parameters of both HD 98618 and 18 Sco are 
very similar (within a few percent) to the host star of our solar system, 
including the likelihood of hosting a terrestrial planet within their habitable zone. 
We suggest that these stars should be given top 
priority in exoplanet and SETI surveys.

\end{abstract}


\section{Introduction}

The most important question for extrasolar terrestrial planet surveys is:
{\it which nearby stars should be searched for Earth-like planets?} (Seager 2003).
We are living proof that stars like our Sun can host habitable planets. 
Thus, stars identical to the Sun (solar twins) may be the best targets 
for future space missions that will probe Earth-like planets for 
signs of chemicals that are only produced when life is present.

Since solar twins are very similar (ideally, identical) 
to the Sun in their fundamental parameters (e.g., temperature, 
luminosity, metallicity, mass, age),
their atmospheric parameters should be very well known,
which makes them extremely important for the calibration of 
fundamental observables  in astrophysics, 
such as the transformation from color to temperature.
In particular, we do not yet know accurately the solar {\it(B-V)} color, 
for which the early literature (1957-1982) shows values
from 0.615 to 0.685 (Cayrel de Strobel 1996). Modern values
(1994-2006) still show a large spread from
a ``blue'' ($\approx$0.60-0.63: Straizys \& Valiauga 1994;
Taylor 1994; Colina, Bohlin \& Castelli 1996;  Sekiguchi \& Fukugita 2000;
Ram\'{\i}rez \& Mel\'endez 2005) to a ``yellow'' ($\approx$0.64:  
Straizys \& Valiauga 1994; Holmberg et al. 2006) to a 
``red'' ($\approx$0.65-0.68: Gray 1995; 
Bessell et al. 1998) {\it (B-V)}$_\odot$ color. 

Twins could be useful in the absolute flux calibration of 
terrestrial and space photometric systems. For example, the recent
absolute calibration of the Infrared Array Camera (IRAC) on the
Spitzer Space Telescope (Reach et al. 2005) yielded 
inconsistent (by up to 7\%) results for A and K stars.
With a photometric stability of just over 1\%, IRAC is capable of 
high-precision photometry; however, the accuracy of IRAC data
is limited by errors in its absolute flux calibration.

Unusual variations in solar activity seem 
to be related to climate changes,
as may be the case for the ``Maunder minimum'' (Maunder 1890;
Eddy 1976), a period of extremely low solar activity that is
related to a ``little ice age'' (Eddy 1976). Thus,
it is of the greatest importance to monitor the activity of stars 
very similar to the Sun. 

The identification of stars closely resembling our host star will be 
of great benefit in answering the long-standing question of the uniqueness 
of the Sun (e.g., Gustafsson 1998), allowing us to fairly determine
whether the Sun is anomalous or not.
Cayrel de Strobel (1996) reviewed the efforts made in the
first 25 years of the search for solar twins and concluded
that none of the large number of stars analyzed is a ``perfect good
solar twin.'' Although a perfect solar twin has yet to be found, 
Porto de Mello \& da Silva (1997) found that 18 Sco (HD 146233) is 
``the closest ever solar twin.'' Soubiran \& Triaud (2004)
show that 18 Sco is the top solar analog in a sample of
about 200 G dwarfs. Very recently, King, Boesgaard \& Schuler (2005)
analyzed four twin candidates and suggested 
that HD 143436 is a solar twin as good as 18 Sco.

Here we show that HD 98618 and 18 Sco are the closest 
solar twins, both with fundamental properties very similar to our Sun.

\section{Data and Analysis}

The candidate solar twins were selected based on their 
blue-optical-red-infrared colors (Ram\'{\i}rez \& Mel\'endez 2005), 
chromospheric activity, v sin $i$, Hipparcos parallaxes
and from previous spectroscopic analysis in the literature
(e.g., Valenti \& Fischer 2005).
The solar twin sample is being observed using HIRES
at the Keck I Telescope during other observing programs.
We are using the upgraded HIRES spectrograph with the highest 
resolving power (R $\approx 10^5$), covering 
$\approx$0.4-0.8 $\mu$m on the ``blue,'' ``green,''
and ``red'' chips. 

So far, 16 candidates (listed in Table 1) have been observed in 2005 and 2006.
Each star was observed for 1-3 minutes in order to achieve a 
S/N of at least $\approx$300 per resolution element 
(or about 200 pixel$^{-1}$) at 671 nm. For most stars, we fulfilled
this requirement. During each run, a solar reflected spectrum was obtained 
by observing either Ceres or Vesta. The data were reduced employing the latest version of 
MAKEE (T. Barlow 2005, private communication). 
Further data reductions (Doppler correction and continuum normalization)
were performed with IRAF. Sample spectra of the Sun,
18 Sco, HD 98618, and HD 143436 are shown in Figure 1.

Orders falling on the ``green'' chip, which in our setup covers $\approx$0.53-0.69 $\mu$m, 
were scrutinized for the best lines available for analysis. We selected
lines of moderate strength (typically with line depths of 0.2-0.6,
but weaker and stronger lines were also included), 
unblended, relatively symmetric and free
of telluric contamination. Lines too close to the edges of the orders 
were discarded because of the greater uncertainty in fitting the continuum in these regions.

We employed two basic criteria to determine how close a star is
to the Sun. The first criterion assessed the similarity between spectral lines in 
the sample star relative to the Sun for different chemical elements. For each atomic line, 
we determined 
the relative difference in line depth $\delta d_r$ employing the 3 closest pixels 
to the line center. We critically selected 206 lines of 10 ions, 
with the largest proportion (70) due 
to \ion{Fe}{1}. A median $<\delta d_r>$ was determined
for each ion and errors were estimated.
A perfect solar twin should have $<\delta d_r>$ = 0 for
lines of all the chemical elements,
and although none was found, 18 Sco and HD 98618 showed the $<\delta d_r>$
values closest to zero (i.e., solar) among the different chemical elements analyzed.

The second criterion allowed us to identify how close a star is
in temperature to the Sun. The principle behind this 
is the excitation equilibrium of \ion{Fe}{1} lines,
which are very sensitive to effective temperature.
For a star identical to the Sun, the slope of a 
plot of $\delta d_r$(\ion{Fe}{1}) versus excitation potential $\chi_{\rm exc}$ 
should be zero. 
In addition to the \ion{Fe}{1} lines used in our first criterion,
we selected four lines around 525 nm from the ``blue'' chip, 
three of them close to $\chi_{\rm exc}$ = 0 eV and
hence allowing a better determination of the d($\delta d_r$)/d($\chi_{\rm exc}$) slope.

The results of the two criteria are shown in Table 1 and Figure 2, 
where it can be seen that 
18 Sco is the best solar twin out of our candidates, with HD 98618 also
appearing very close to the Sun.  Figure 2 also shows that,
according to our  criteria, HD 143436 is no match
for 18 Sco, in contrast to the suggestion (King et al. 2005) that they may be
equivalent. In addition, we estimated the relative depths of 
the Li feature at 6708 \AA. HD 98618  and 18 Sco have similar Li profiles, which are 
both deeper than solar yet much closer to the Sun's than the Li profile of HD 143436.
Hence, also considering the Li abundance, HD 143436 is not a good solar twin.
The spectroscopic log $g$ (= 4.28) determined by King et al. (2005) 
is considerably lower than the trigonometric log $g$ from Hipparcos 
parallaxes ($\approx$4.46, using the \teff and mass given by them), 
which suggests errors in their analysis. Since the spectroscopic 
log $g$ is too low, the spectroscopic \teff given by King 
et al. is probably also underestimated.

It is important to note that our analysis is strictly differential and independent
of model atmospheres, with both the sample 
and reflected solar spectrum (Vesta or Ceres) obtained employing the 
same HIRES setting and during the same run and the data reduction and
continuum normalization performed in the same way. 
On the other hand, King et al. (2005) 
used a lunar spectrum obtained several years earlier, and their
observations are of lower resolving power (R $\approx$ 45,000).
The set of lines used is also different; while King et al. cover 647-675 nm,
our \ion{Fe}{1} lines cover 522-675 nm.
Note that part of the region used by them has weak telluric contamination.

\section{Detailed Analysis of 18 Scorpii and HD 98618}

The top twin candidates, 18 Sco and HD 98618, were 
analyzed in more detail to quantify their likeness to
the Sun by performing a differential spectroscopic equilibrium
analysis with
respect to the solar spectrum. The close proximity of our candidates 
to the Sun warrants reliable {\it relative} chemical and physical parameters.

The equivalent widths of 39 \ion{Fe}{1} and 14 \ion{Fe}{2} lines with 
clean profiles and well-defined continuum regions were measured 
in the region 439-675 nm. The $gf$-values of \ion{Fe}{1} and 
\ion{Fe}{2} lines were taken from the Oxford (e.g., Blackwell,
Lynas-Gray \& Smith 1995) and Hannover groups (e.g., Bard \& Kock 1994)
and from Mel\'endez et al. (2006), respectively.
Oxygen and lithium abundances were obtained
from the \ion{O}{1} triplet at 777 nm 
and the \ion{Li}{1} profile at 670.7 nm.
The interaction broadening constants $C_6$
were obtained from the collision broadening cross sections 
given by Barklem, Piskunov \& O'Mara (2000)  and
Barklem \& Aspelund-Johansson (2005).

The LTE calculations were performed with the 2002 version of MOOG (Sneden 1973),
employing Kurucz overshooting model atmospheres
(Castelli et al. 1997).
The microturbulence v$_t$  was determined by 
requiring no dependence of the iron abundance from \ion{Fe}{1} lines (A$_{FeI}$) 
against reduced equivalent width ($W_\lambda/\lambda$).

The zero point of the excitation and ionization equilibrium
was defined by the analysis of the solar \ion{Fe}{1} and
\ion{Fe}{2} lines:
(dA$_{FeI}$/d$\chi_{\rm exc}$)$_\odot$ and
($\Delta_{FeII - FeI}$)$_\odot$ (= A$_{FeII}$ - A$_{FeI}$).
For the Sun we adopted the atmospheric 
parameters (\tsin, log $g$, [Fe/H] = 5777 K, 4.44, 0.0), which was
also adopted as a first guess for the twin candidates.
Then the parameters of the sample stars were 
iteratively changed in ($\Delta$\tsin, $\Delta$log $g$, $\Delta$[Fe/H]) 
until we simultaneously achieved {\it relative} excitation 
and ionization equilibrium: \\

(dA$_{FeI}$/d$\chi_{\rm exc}$)$_*$  = (dA$_{FeI}$/d$\chi_{\rm exc}$)$_\odot$ \\

($\Delta_{FeII - FeI}$)$_*$ = ($\Delta_{FeII - FeI}$)$_\odot$ \\

The stellar parameters so determined were used to obtain
stellar masses and ages from Y$^2$ isochrones 
(Demarque et al. 2004). Once the mass  was
estimated, a spectroscopic luminosity was obtained from 
log ($L$/$L_\odot$) = log ($M/M_\odot$) - log($g/g_\odot$) + 4 log (\tsin/\tsin$_\odot$).

We employed Hipparcos parallaxes to check both the surface 
gravities and luminosities obtained in the present spectroscopic analysis.
The trigonometric gravity (log $g_{\rm Hip}$) and 
stellar luminosity ($L_{\rm Hip}$) were calculated employing Hipparcos 
parallaxes, V, M${_{bol}}_\odot$ = 4.736
(Bessell et al. 1998), and the bolometric correction\\

BC = -1.6240 + 4.5066 $\theta$ -3.12936 $\theta^2$ ($\sigma$ = 0.02 mag)\\

where $\theta$ = (5040 K)/\tsin. BC was obtained from a fit
to solar metallicity dwarfs from Alonso et al. (1995) and
Blackwell \& Lynas-Gray (1998); the relation above is valid from about
4200 to 7000 K. 

The spectroscopic and Hipparcos-based results are shown
in Table 2, where one can see the excellent agreement in both
log $g$ and luminosity, which lends confidence to the results
obtained in the present work. A comparison with previous
differential spectroscopic analyses is shown in Table 3.
All agree that the atmospheric parameters
of 18 Sco and HD 98618 are very similar to the Sun,
especially in \teff (1.1\% or better).
The differences in log $g$ and [Fe/H] are just
within a few times 0.01 dex, almost the same as the typical
uncertainties. Although within the uncertainties all these works
agree with each other, the trigonometric gravities from 
Hipparcos parallaxes (Table 2) favor our results.

Rotation periods and stellar ages can be obtained from measurements
of stellar activity. Table 2 shows the relative 
chromospheric activity indices (log $R'_{\rm HK}$), 
chromospheric ages and rotation periods estimated
by Wright et al. (2004). A better rotation
period for HD 146233 is given by Frick et al. (2004), who
analyzed time series of chromospheric activity.

The rotation periods can be used to estimate a rotational age (Barry 1988).
The relation between age $t$ and rotation period $P$
also depends on mass [log $P$ = a log $t$ + b log($M/M_\odot$) + c; 
Kawaler 1989), but since our stars have about 1$M_\odot$,
this relation simplifies to $P$ $\propto$ $t^a$. Barry (1988) finds
that $a$ $\approx$ 1/e, and using a mean solar rotation period 
(equator to latitude 20$\degr$) $P_\odot$ $\approx$ 25.3 days (Howard 1984) 
and a solar main-sequence age of $t_\odot$ = 4.51 Gyr (Sackmann, Boothroyd \&
Kraemer 1993), we find $t$ = ($P$/14.5)$^e$, where $P$ is in days and $t$ is in Gyr. 
Using this relation, we estimated rotational ages (Table 2) adopting $P$ from
Frick et al. (2004) and Wright et al. (2004) for 18 Sco and HD 98618, respectively.

\section{Discussion and Conclusions}

As shown in Table 2, 18 Sco and HD 98618 are both very similar to the Sun, 
with the surface gravity being just 0.01 dex higher than
solar, the mass above solar by 2\%, \teff hotter by 0.7\% and 1.1 \%, 
metallicity higher by 0.02 and 0.05 dex, and luminosity about 3\% and 6 \% higher,
respectively. Both stars belong to the disk population, and, as the Sun,
are orbiting the center of our Galaxy at $\approx$8 kpc (Nordstr\"om et al. 2004).

Both twin candidates seem to be rotating slightly faster than the Sun 
and are about 10\% younger than the Sun. The higher Li abundance in 
18 Sco and HD 98618 is perhaps due to different amounts of pre-main-sequence 
depletion (see, e.g., Ventura et al. 1998). 
Since 18 Sco and HD 98618 are about 4 Gyr old, hypothetical
terrestrial planets may have had enough time to develop some kind 
of complex life, assuming the time-scale for formation of
complex life is similar to Earth's. Thus, we encourage SETI programs to give top
priority to these stars.

Thanks to the long-term effort of the California and Carnegie 
Planet Search Project (see, e.g.,
Marcy et al. 2005), we know that there are not any hot Jupiters around
18 Sco or HD 98618. According to Marcy et al. (2005) 
``among the nearest FGK main-sequence stars (d $<$ 40 pc) 
the yet undiscovered giant planets typically reside beyond 1 AU 
because giant planets within 1 AU have already been found.''
Furthermore, Marcy et al. estimate that undetected Jupiter-mass 
planets orbiting nearby FGK stars ($<$30 pc) may reside in orbits beyond 3 AU 
or have masses less than $M_{\rm Jup}$.
Since the distance at which an Earth-size planet would receive the same amount of 
light from its star as Earth receives from the Sun is ($L/L_\odot$)$^{1/2}$AU 
and the luminosities of the twin candidates are very close to solar, 
the habitable zones of 18 Sco and HD 98618 should be very close
to 1 AU. If further radial velocity observations show that the 
inner region around 18 Sco and HD 98618
is free of giant planets, then these stars
have the potential to host terrestrial planets around
their habitable zones. We encourage long-term radial velocity 
monitoring of these stars, which could reveal within a few years how far out gaseous 
giant planets, if any, reside within the exoplanetary systems these stars may host.

{\it Note added in proof}.- After this Letter was accepted for publication, M. C. Turnbull
brought it to our attention that HD 98618 is included in the Catalog
of Nearby Habitable Stellar Systems (M. C. Turnbull \& J. C. Tarter,
ApJS, 145, 181; ApJS, 149, 423 [2003]) for SETI searches with the Allen
Telescope Array. It is ranked No. 106 in terms of its location on the
H-R diagram, and considering its distance merits its final ranking is
No. 504. The present work suggests that HD 98618 should be given
a higher ranking.

\begin{deluxetable}{lrrrrr}
\tablecolumns{5}
\scriptsize
\tablecaption{Slope d($\delta d_r$)/d($\chi_{\rm exc}$) for
\ion{Fe}{1}, and Relative Differences in Line Depth $<\delta d_r>$
for \ion{Fe}{1}, \ion{V}{1}, and \ion{Ca}{1}}
\tablehead{
\colhead{Star} &
\colhead{Slope}  &
\colhead{\ion{Fe}{1}}  &
\colhead{\ion{V}{1}}  &
\colhead{\ion{Ca}{1}}  
}
\startdata
HD 9986   &  $-$0.0120  &  $-$0.042  &  $-$0.100  &  $-$0.019 \\  
HD 32963  &  $-$0.0075  &  $-$0.112  &  $-$0.301  &  $-$0.080 \\ 
HD 33636  &   0.0048    &   0.220    &   0.445    &   0.116 \\ 
HD 45184  &  $-$0.0063  &   0.010    &   0.098    &   0.004 \\  
HD 56124  &   0.0010    &   0.022    &   0.145    &   0.010 \\ 
HD 71334  &  $-$0.0120  &   0.001    &   0.079    &  $-$0.027 \\  
HD 71148  &  $-$0.0047  &   0.037    &   0.145    &   0.021 \\  
HD 71881  &   0.0120    &   0.077    &   0.203    &   0.045 \\ 
HD 98618  &  $-$0.0030  &  $-$0.003  &  $-$0.016  &   0.002 \\ 
HD 112257 &   0.0046    &  $-$0.047  &  $-$0.292  &  $-$0.070 \\  
HD 143436 &  $-$0.0064  &   0.008    &   0.078    &   0.011 \\  
HD 146233 &  $-$0.0007  &  $-$0.001  &   0.024    &  $-$0.004 \\  
HD 147044 &   0.0040    &   0.090    &   0.270    &   0.055 \\  
HD 159222 &  $-$0.0160  &  $-$0.017  &   0.000    &   0.013 \\ 
HD 186427 &  $-$0.0062  &  $-$0.079  &  $-$0.213  &  $-$0.053 \\  
HD 187123 &  $-$0.0120  &  $-$0.081  &  $-$0.153  &  $-$0.052 \\  
\enddata
\label{sample}
\end{deluxetable}

\begin{deluxetable}{lccccc}
\tablewidth{0pt}
\tablecolumns{3}
\scriptsize
\tablecaption{Fundamental Parameters}
\tablehead{
\colhead{Parameter (Star -- Sun)}   &
\colhead{18 Sco} &
\colhead{HD 98618} &
}
\startdata
$\Delta$ v$_t$ (km s$^{-1}$)      &    +0.08$\pm$0.15    & +0.09$\pm$0.15  \\
$\Delta$ \teff (K)                 &   +40 $\pm$ 30       &  +66 $\pm$ 30   \\
$\Delta$ log $g_{\rm spec}$ (dex)  &  +0.01 $\pm$ 0.04    &  +0.01$\pm$ 0.04 \\
$\Delta$ log $g_{\rm Hip}$ (dex)   &  +0.01 $\pm$ 0.02    &  +0.01$\pm$ 0.03 \\
{\bf $\Delta$ log $g_{\rm adopted}$ (dex)} & {\bf +0.01 $\pm$ 0.02} &  {\bf +0.01 $\pm$ 0.03}\\
$\Delta$ L$_{\rm spec}$ (L$_\odot$) &  +0.02 $\pm$ 0.06    &  +0.04$ \pm$ 0.06 \\
$\Delta$ L$_{\rm Hip}$  (L$_\odot$) &  +0.03 $\pm$ 0.03    &  +0.08$ \pm$ 0.07 \\
{\bf $\Delta$ L$_{\rm adopted}$ (L$_\odot$)} & {\bf +0.03 $\pm$ 0.02} &  {\bf +0.06$\pm$ 0.05}\\
{[Fe/H] (dex)}                    &   +0.02 $\pm$ 0.03  & +0.05$\pm$ 0.03 \\
{[O/H] (dex)}                     &   $-$0.03 $\pm$ 0.05  &  0.00 $\pm$ 0.04 \\
{[Li/H] (dex)}                    &   +0.53 $\pm$ 0.09   & +0.47 $\pm$ 0.09 \\
$\Delta$ mass ($M_{\odot}$)      &   +0.02 $\pm$ 0.03  & +0.02 $\pm$ 0.03 \\
$\Delta$ age$_{\rm isochro}$ (Gyr) &     $-$0.8 $\pm$ 1.5   &   $-$1.1 $\pm$ 1.5 \\
$\Delta$ age$_{\rm chromos}$ (Gyr) &       $-$0.3$^1$     &     +0.7$^1$   \\
$\Delta$ age$_{\rm rotation}$ (Gyr) &      $-$1.1         &     $-$0.4     \\
{\bf $\Delta$ age$_{\rm adopted}$ (Gyr)} &  {\bf $-$0.7 $\pm$ 0.4} &  {\bf $-$0.3 $\pm$ 0.9} \\
$\Delta$ rotation period (days)    &   $-$2.5$^2$, -1$^1$   &    $-$1$^1$  \\
$\Delta$ log{$R'_{\rm HK}$} (dex)   &        0.0$^1$     &       $-$0.05$^1$  \\
$\Delta ${$M_{V}$} (mag)           & $-$0.04 $\pm$ 0.04    & $-$0.09 $\pm$ 0.07 \\
{\it  B-V}                     &         0.65         &     0.64        \\
{Distance}{(pc)}                &         14.0        &     38.7        \\
\enddata
\tablerefs{(1) Wright et al. 2004; (2) Frick et al. 2004}
\label{fundamental}
\end{deluxetable}

\begin{deluxetable}{ccl}
\tablewidth{0pt}
\tablecolumns{4}
\scriptsize
\tablecaption{Comparison of Relative Spectroscopic 
Parameters $\Delta$(\tsin/log $g$/[Fe/H])}
\tablehead{
\colhead{18 Sco} &
\colhead{HD 98618}  &
\colhead{Ref.}   \\
\colhead{(K/dex/dex)} &
\colhead{(K/dex/dex)}  &
\colhead{}   
}
\startdata
 +40/+0.01/+0.02  &  +66/+0.01/+0.05 & 1\\
 +12/+0.05/+0.05  &          & 2\\
 +58/+0.03/+0.03  &          & 3 \\
+14/$-$0.03/+0.03   &  +35/$-$0.02/+0.03  & 4\\
\enddata
\tablerefs{(1) This work; (2) Porto de Mello \& da Silva (1997); (3) Luck \& Heiter (2005);
(4) Valenti \& Fischer (2005)}
\label{comparison}
\end{deluxetable}

\begin{figure}
\includegraphics[scale=0.8,angle=0]{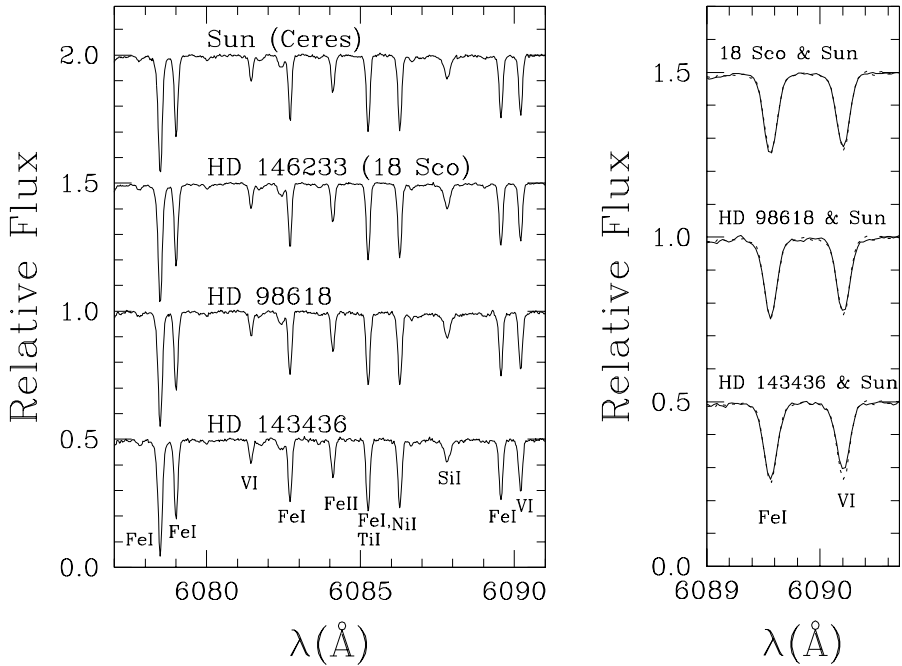}
\caption{{\it Left}, sample spectra of the Sun and the candidate solar twins
18 Sco, HD 98618, and HD 143436; {\it right}, comparison of the
solar-twin candidates (solid lines) with the Sun (dotted line)
around 6090 {\AA}. Since \ion{V}{1} is very sensitive to \tsin, 
the weaker  \ion{V}{1} line in  HD 143436 suggests that
this star has a \teff higher than solar, while
both 18 Sco and HD 98618 are very similar to the Sun.}
\label{spectra} 
\end{figure}

\begin{figure}
\includegraphics[scale=.50,angle=0]{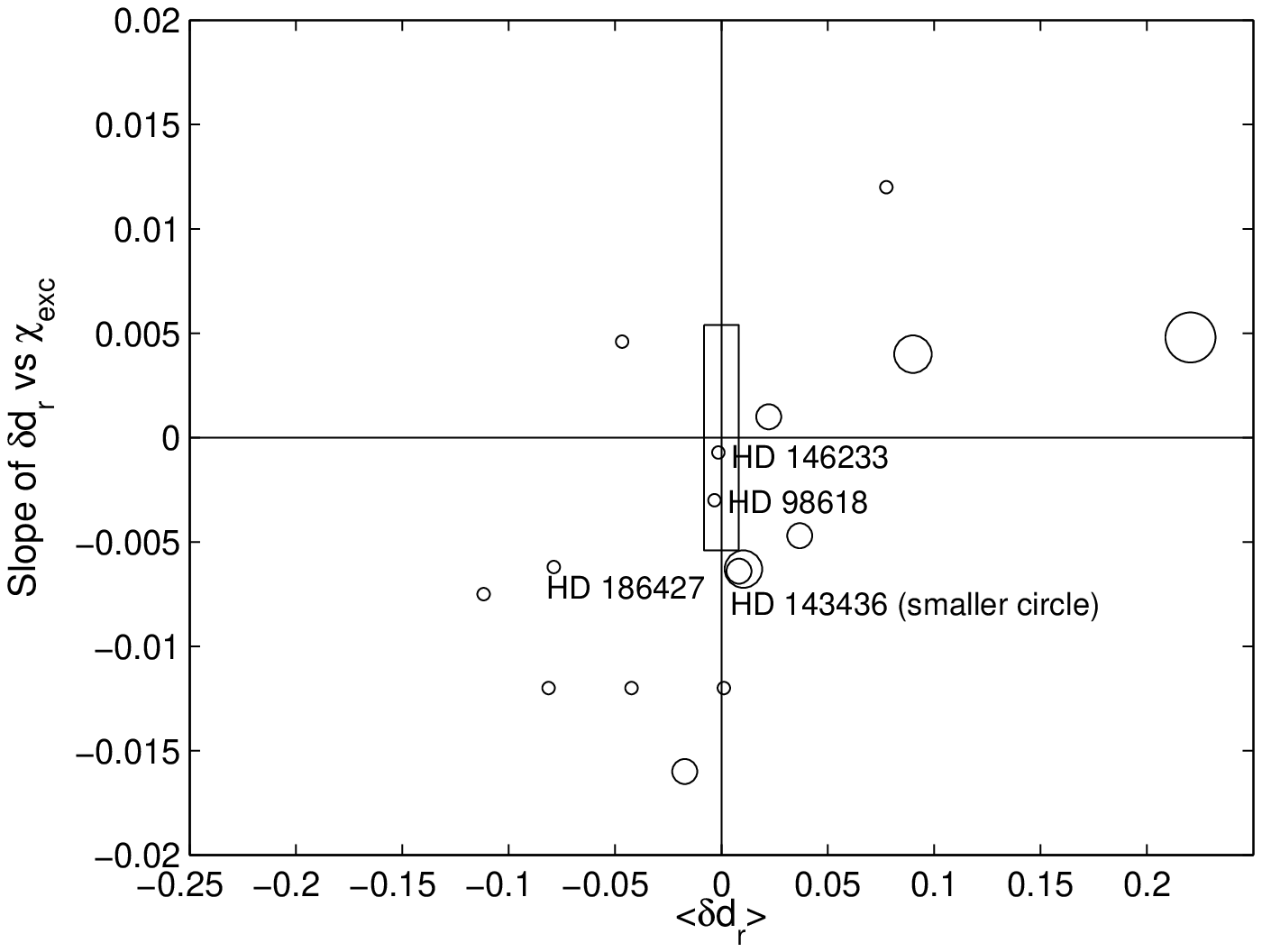} 
\includegraphics[scale=.50,angle=0]{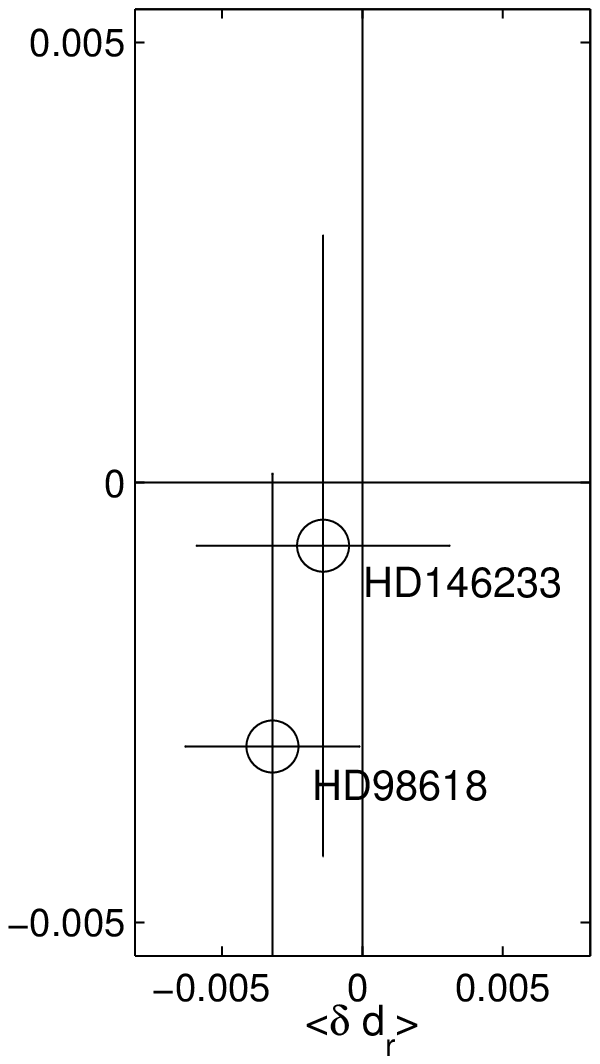}
\caption{Slope d($\delta d_r$)/d($\chi_{\rm exc}$)  vs. the median 
relative difference in line depth $<\delta d_r>$ for \ion{Fe}{1} lines ($\S$2).
A perfect solar twin should be at (0,0). 
The size of the circles is proportional to the \ion{Li}{1} line depth at 6708 {\AA},
with the smallest circles representing closer agreement to the solar \ion{Li}{1} profile.
{\it Left}, the whole sample, with the inner rectangle 
representing the typical $\pm1 \sigma$ error bars; {\it right}, a zoom to the
inner rectangle of the left panel. The two best solar twin candidates,
18 Sco (HD 146233) and HD 98618, are labeled; HD 143436 
(King et al. 2005) and HD 186427 (one of the most promising 
twin candidates in the early literature) are shown as well.}
\label{spectra} 
\end{figure}

\end{document}